\documentclass[twocolumn,secnumarabic,amssymb, nobibnotes, aps, prd]{revtex4}
\usepackage{graphicx}

\begin{document}
\title{Could two degenerate energy states be observed for a superconducting ring at $\Phi_{0}$/2?}
\author{A.A. Burlakov, V.L. Gurtovoi ,  S.V. Dubonos, A.V. Nikulov, and V.A. Tulin}
%\email[]{nikulov@ipmt-hpm.ac.ru}
\affiliation{Institute of Microelectronics Technology and High Purity Materials, Russian Academy of Sciences, 142432 Chernogolovka, Moscow District, RUSSIA.} %nikulov@ipmt-hpm.ac.ru
%\date{}
\begin{abstract}The Little-Parks oscillations of the resistance and the quantum oscillations of the rectified voltage observed for asymmetric superconducting Al rings give experimental evidence of two degenerate energy states at the magnetic flux $\Phi_{0}$/2. The quantum oscillations of the critical current as a function of magnetic field have also been measured. On the one hand, these oscillations confirm that the quantum oscillations of the rectified voltage are a consequence of periodical dependence of the asymmetry of the current-voltage curves and, on the other hand, comparison of the oscillations with Little-Parks measurements results in contradiction.  \end{abstract}

\maketitle

\narrowtext

One of the most intriguing problems of mesoscopic physics is a possibility of superposition of quantum macroscopic states \cite{Legget02}. 
It is especially urgent because of the aspiration for realization of the idea of the quantum computation \cite{QuanCom}. The quantum computations has 
grown from both the paradoxical nature of the quantum principles and the contradiction between quantum mechanics 
and local realism \cite{Steane} revealed by A. Einstein, B. Podolsky, and N. Rosen in 1935 \cite{EPR} and put into the inequalities by J.S. 
Bell in 1964 \cite{Bell}. The experimental evidence of violation of local realistic predictions has been obtained for the present only on 
the level of  particles, first of all photons \cite{nonlocal}. On the mesoscopic level, quantum mechanics contradicts not only to local 
but also to macroscopic realism \cite{Legget85}. According to the principle of realism {\it a macroscopic system with two or more macroscopically 
distinct states available to it will at all times be in one or the other of these states} \cite{Legget85}, i.e. quantum superposition is not 
possible. A. J. Leggett and A. Garg consider this contradiction on the example of rf SQUID, i.e. superconducting loop interrupted by Josephson 
junction. This consideration is based on the assumption on two permitted state in such loop with half quantum $\Phi_{0}= \pi \hbar/e$ of magnetic 
flux $\Phi = (n+0.5)\Phi_{0}$. The authors \cite{Legget85} do not doubt that the two states exist and any observation will find a value 
corresponding to one of them. But nobody must be sure of anything in the quantum world till unambiguous experimental evidence. A. Einstein, 
B. Podolsky, and N. Rosen were sure \cite{EPR} that a process of measurement carried out on a one system can not affect other system in any way. 
But experimental results \cite{nonlocal} have shown that it can and this phenomenon is called now Einstein - Podolsky- Rosen correlation. Some 
experts of quantum physics doubt about the reality even of the moon o
n the night sky \cite{Mermin}.

According to the universally recognized explanation \cite{tink75} the numerous observation of the Little-Parks oscillations \cite{Little62} of 
resistance $\Delta R(\Phi/\Phi_{0})$ of superconducting loop \cite{Mosh92} prove quantization of velocity circulation
$$\oint_{l} dl v = \frac{2\pi \hbar}{m} (n -\frac{\Phi}{\Phi_{0}}) \eqno{(1)}$$
of superconducting pairs and that the permitted state with minimum energy has overwhelming probability even at 
$T \approx T_{c}$. This explanation assumes two permitted states, $n$ and $n+1$, with minimum energy at $\Phi = (n+0.5)\Phi_{0}$ 
but the Little-Parks oscillations can not prove this since the resistance $\Delta R(\Phi/\Phi_{0}) \propto v^{2} \propto (n -\Phi/\Phi_{0})^{2}$ 
\cite{tink75} has the same value, $\propto (-1/2)^{2}$ and $\propto (1/2)^{2}$, at $n$ and $n+1$.
\begin{figure}
\includegraphics{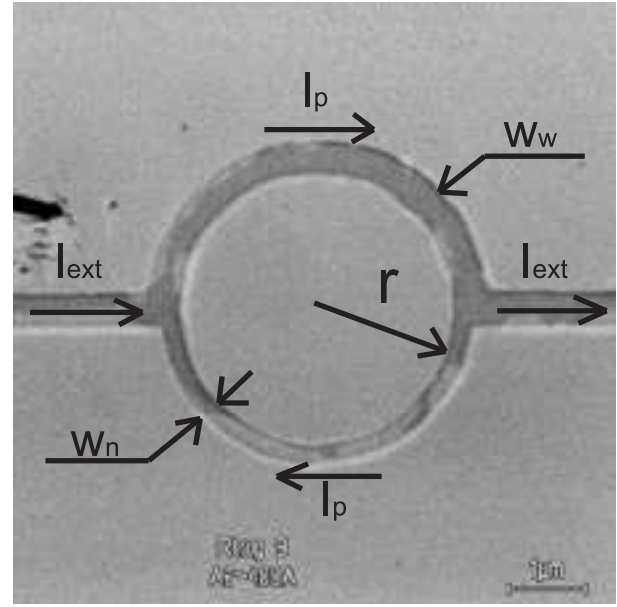}
\caption{\label{fig:epsart} An SEM image of the asymmetric Al ring with radius $r = 2 \ \mu m$ and semi-ring width $w_{n} = 0.2 \ \mu m$, $w_{w} = 0.4 \ \mu m$. 
Directions accepted as positive of the external current $I_{ext}$ and the persistent current $I_{p}$ are shown by arrows.}
\end{figure}
At the geometry, Fig.1, used for the observation of the Little-Parks oscillations \cite{Mosh92}, the velocity, $v_{n}$, $v_{w}$ of superconducting pairs in the loop half is determined by an external current $I_{ext} = I_{n}+I_{w} = s_{n}2en_{sn}v_{n} + s_{w}2en_{sw}v_{w}$ and the quantization (1): $l_{n}v_{n} - l_{w}v_{w} = (2\pi \hbar/m) (n -\Phi/\Phi_{0})$. The velocity will mount to the critical value $v_{c} = \hbar /m\xi (T)\surd 3 $ \cite{tink75} in the loop with equal half length $l_{n} = l_{w} = l/2$ and an equal density of superconducting pairs $n_{s} = n_{sn} = n_{sw}$ at the external current $I_{c+}, I_{c-} = 2en_{s}(s_{n}v_{c} + s_{w}v_{c} - s_{w}(2\hbar /mr)|n - \Phi/\Phi_{0}| = I_{c} - |I_{p}|(s_{n} + s_{w})/s_{n}$ when $|v_{n}| = v_{c}$ or $I_{c+}, I_{c-} = I_{c} - |I_{p}|(s_{n} + s_{w})/s_{w}$ when $|v_{w}| = v_{c}$. Here $I_{c} = 2en_{s}(s_{n} + s_{w})v_{c}$ is the critical current which should be measured at $n -\Phi/\Phi_{0} = 0$; $I_{p} = 2en_{s}[2s_{n}s_{w}/(s_{n}+s_{w})](\hbar/mr) (n -\Phi/\Phi_{0})$ is the persistent current, i.e. the circular direct current flowing in the loop halves $I_{p} = I_{n} = -I_{w}$ without the external current $I_{ext} = I_{n}+I_{w} = 0$; $r$ is the radius of the round loop, i.e. ring, Fig.1, $l = 2\pi r$; $I_{c+}$ and $I_{c-}$ are the critical current measured in opposite directions of $I_{ext}$. Here and below the positive values correspond the right-left direction of $I_{ext},  I_{n}, I_{w}$ and the clockwise one of $I_{p}$, Fig.1.

Measurements of the critical current $I_{c+}, I_{c-}$ of asymmetric superconducting loop with different sections $s_{n} \neq s_{w}$ of the semi-rings may be used for an investigation of the problem of two permitted states at $\Phi = (n+0.5)\Phi_{0}$ since the $I_{c+}, I_{c-}$ values depend not only on value but also on sign of $I_{p} \propto n -\Phi/\Phi_{0}$. The anisotropy of the critical current $I_{c,an} = I_{c+} - I_{c-}$ should be proportional to the persistent current $I_{c,an} = I_{p}(s_{w}/s_{n} - s_{n}/s_{w})$: the velocity increases mounts the critical value in the wide half $|v_{w}| = v_{c}$ when $I_{ext}$ and $I_{p}$ have the same sign and in the narrow one $|v_{n}| = v_{c}$ when they have the opposite sign, Fig.1. The periodical change of the critical current because of the quantization (1) can be appreciable when the amplitude of the persistent current $I_{p,A} = 2en_{s}[s_{n}s_{w}/(s_{n}+s_{w})](\hbar/mr)$ is not too small in comparison with the critical current $I_{c} = 2en_{s}(s_{n} + s_{w})\hbar /m\xi (T)\surd 3 $. The relation $I_{p,A}/I_{c} \simeq \xi (T)\surd 3/4r$ can be enough high at radius $r \approx 1 \mu m$ available for modern microtechnology at using superconductor with large coherence length $\xi $, such as aluminum.

We used for the investigation asymmetric aluminum rings with radius $r = 2 \mu m$, thickness $d = 40-70 \ nm$, semi-ring width $w_{n} = 0.2 \ \mu m$, $w_{w} = 0.4; 0.3; 0.25 \ \mu m$, Fig.1 and also symmetrical rings with $w_{n} = w_{w} = 0.4 \ \mu m$. This nano-structures were fabricated by thermally evaporated on oxidized Si substrates, e-beam lithography and lift-off process. Their sheet resistance was $0.2 \div 0.5 \ \Omega /\diamond $ at 4.2 K, the resistance ratio $R(300K)/R(4.2K)=2.5 \div 3.5$, and critical temperature was $T_{c} = 1.24 \div 1.27 \ K$. The coherence length of the like aluminum structures $\xi (T) = \xi (0)(1 - T/T_{c})^{-1/2} \approx 200 \ nm (1 - T/T_{c})^{-1/2}$, the London penetration depth $\lambda_{L} (T) = \lambda_{L} (0)(1 - T/T_{c})^{-1/2} \approx 50 \ nm (1 - T/T_{c})^{-1/2}$ and the thermodynamic critical field $H_{c}(T) = H_{c}(0)(1 - T/T_{c}) \approx 0.01 \ T (1 - T/T_{c})$.

\begin{figure}
\includegraphics{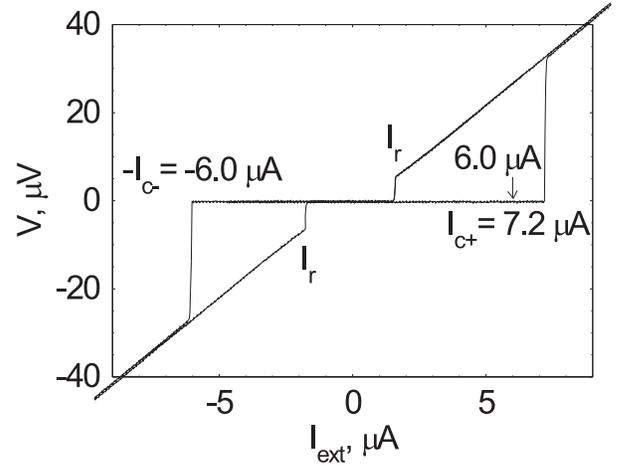}
\caption{\label{fig:epsart} The current-voltage curve measured on an asymmetric aluminum ring with $r = 2 \mu m$, $w_{n} = 0.2 \ \mu m$, $w_{w} = 0.4 \ \mu m$, $T_{c} = 1.243 \ K$ at $T = 1.235 \ K$ }
\end{figure}

Our measurements in the temperature region $T = (0.95 \div 1.0)T_{c}$ have shown that the current-voltage curves are strongly irreversible, Fig.2, at $T \leq 0.995T_{c}$ and the experimental values of the critical current $I_{c}/(s_{n} + s_{w}) = j_{c} = 10^{11} \ A/m^{2} (1 - T/T_{c})^{3/2}$ corresponding to the jump in the normal state are close to the theoretical one $ j_{c} = 2en_{s}(v_{c})v_{c} = H_{c}(T)/3\surd 6 \pi \lambda_{L} (T) = H_{c}(0)/3\surd 6 \pi \lambda_{L} (0) (1 - T/T_{c})^{3/2}$ \cite{tink75}. One should not expect any change of the quantum number $n$ up to the jump in the normal state since the number of the pairs $\pi r(s_{n} + s_{w})n_{s}(v_{c})$ in the ring \cite{Nik2001} remains very big up to $v = v_{c}$ when $n_{s}(v_{c}) = 2n_{s}(v=0)/3$ \cite{tink75}. Therefore the measurement of the critical current $I_{c+}, I_{c-}$ should corresponds single measurement of permitted state $I_{p} \propto n -\Phi/\Phi_{0} = 1/2$ or $-1/2$ at $\Phi = (n+0.5)\Phi_{0}$, Fig.3. Whereas the average value obtained at a multiple measurement should be equal zero $\overline{I_{p}} \propto \overline{n} -\Phi/\Phi_{0} \propto  1/2 + (-1/2) = 0$, Fig.3.

\begin{figure}
\includegraphics{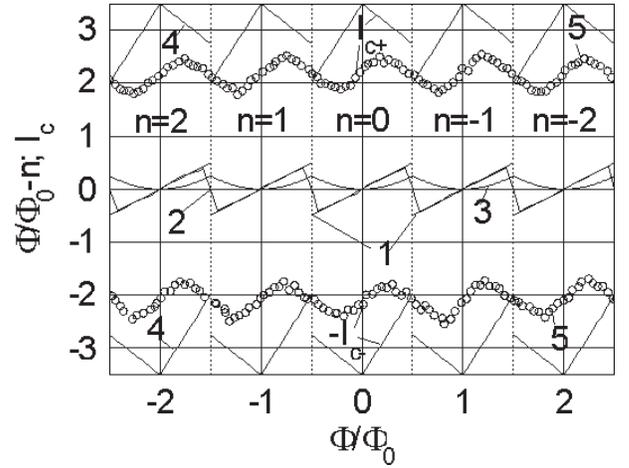}
\caption{\label{fig:epsart} Magnetic dependencies of the permitted velocity $v \propto n -\Phi/\Phi_{0}$ (1), the average equilibrium velocity $\overline{v} \propto \overline{n} -\Phi/\Phi_{0}$ (2), the velocity square $v^{2} \propto (n -\Phi/\Phi_{0})^{2}$ (3) and the critical current (4) expected $I_{c+}, I_{c-} = I_{c} - |I_{p}|(s_{n} + s_{w})/s_{w}(or s_{n})$ for the ring with $s_{w}/s_{n} = 2$,  $I_{c} = 3.5 \ \mu A$, $I_{p} = (n -\Phi/\Phi_{0})  \ \mu A$ (line) and observed on Al loop with  $w_{n} = 0.2 \ \mu m$,  $w_{w} = 0.4 \ \mu m$, $T_{c} = 1.235 \ K$ at  $T = 1.225 \ K$ ($\circ $).}
\end{figure}

We have found that the values of the critical current corresponding to the jump in the normal state change periodically in magnetic field $I_{c+}(\Phi/\Phi_{0}), I_{c-}(\Phi/\Phi_{0})$ whereas the value of the current $I_{r}$ at which the ring reverts in superconducting state, Fig.2, does not depend on magnetic field and the current $I_{ext}$ direction. The periodical dependence of the anysotropy of the critical current $I_{c,an}(\Phi/\Phi_{0})$ corroborate the interpretation of the quantum oscillations of the dc potential difference $V_{dc}(\Phi/\Phi_{0})$ \cite{Dub2003} as a consequence of the rectification of the ac external current $I_{ext}(t)$: $V_{dc} = \Theta ^{-1}\int_{\Theta }dt V(I_{ext}(t))$. The quantum oscillations $V_{dc}(\Phi/\Phi_{0})$ appear when the amplitude $I_{0}$ of the external current $I_{ext}(t) = I_{0}\sin(2\pi ft)$ exceeds minimum from the $I_{c+}, I_{c-}$ values $I_{0} > min(I_{c+}, I_{c-})$, its amplitude $V_{A}$ mounts the maximum at $max(I_{c+}, I_{c-}) > I_{0} > min(I_{c+}, I_{c-})$ and decreases when the $I_{0}$ exceeds maximum from the $I_{c+}, I_{c-}$ values, see Fig.2,6 in \cite{Dub2003}.

\begin{figure}
\includegraphics{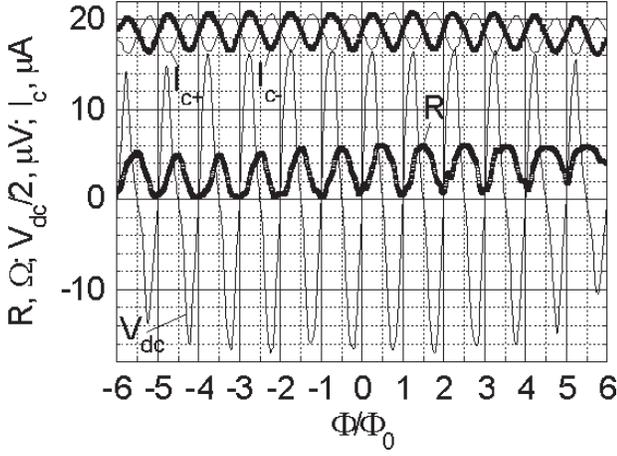}
\caption{\label{fig:epsart} The Little-Parks oscillations $R$ at T = 1.278 K, the quantum oscillations of the rectified voltage $V_{dc}/2$, induced by the ac current with frequency f = 0.5 kHz and amplitude $I_{0} = 20 \ \mu A$ at T = 1.253 K and of the critical current $I_{c+}$, $I_{c-}$ at T = 1.253 K measured on the ring with $w_{n} = 0.2 \ \mu m$,  $w_{w} = 0.3 \ \mu m$ and $T_{c} = 1.288 \ K$.}
\end{figure}

The rectified voltage $V_{dc} = \Theta ^{-1}\int_{\Theta }dt V(I_{ext}(t)) \propto -R\overline{I_{c,an}} = - R\sum_{\Theta f}(I_{c+} - I_{c-})/\Theta f$ at $I_{0} > max(I_{c+}, I_{c-})$ since the current voltage curves $V(I_{ext})$, Fig.2, are symmetrical at $|I_{ext}| < min(I_{c+}, I_{c-})$ and $|I_{ext}| > max(I_{c+}, I_{c-})$. Therefore the measurement of the rectified voltage $V_{dc} \propto -R\overline{I_{c,an}} \propto \overline{I_{p}} \propto \overline{n} -\Phi/\Phi_{0}$ should correspond to multiple measurement of quantum states. The measurements \cite{Dub2003} and of our numerous measurements have shown that all $V_{dc}(\Phi/\Phi_{0})$ dependencies cross zero at $\Phi = n\Phi_{0}$ and $\Phi = (n+0.5)\Phi_{0}$, Fig.4. The results of the measurements of the rectified voltage $V_{dc}(\Phi/\Phi_{0}) \propto \overline{v} \propto \overline{n} -\Phi/\Phi_{0}$ and the Little-Parkse $\Delta R(\Phi/\Phi_{0}) \propto \overline{v^{2}} \propto \overline{(n -\Phi/\Phi_{0})^{2}}$ oscillations made on the same ring, Fig.4, give evidence of two permitted states with the same minimum energy $\propto v^{2} \propto  (n -\Phi/\Phi_{0})^{2} = (1/2)^{2} = (-1/2)^{2}$ at $\Phi = (n+0.5)\Phi_{0}$. There is not possible to explain the quantum oscillations $\Delta R,V_{dc}(\Phi/\Phi_{0})$ and the simultaneous observation of the maximum resistance $\Delta R(\Phi/\Phi_{0}) \propto \overline{v^{2}}$ and the zero voltage $V_{dc}(\Phi/\Phi_{0}) \propto \overline{v} = 0$ without the assumption of two permitted states at $\Phi = (n+0.5)\Phi_{0}$: $\overline{v^{2}} \propto ((1/2)^{2}+ (-1/2)^{2})/2 = 1/4 = max (n -\Phi/\Phi_{0})^{2}$ and $\overline{v} \propto (1/2)+ (-1/2) = 0$, Fig.3.

\begin{figure}
\includegraphics{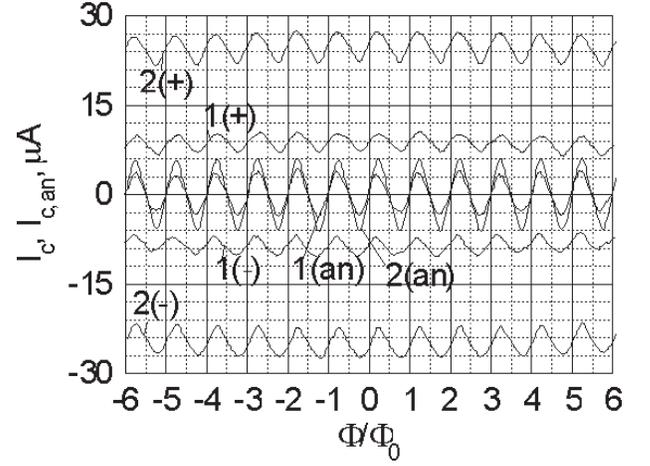}
\caption{\label{fig:epsart} The quantum oscillations of the critical current $I_{c+}$, $I_{c-}$ in opposite directions and its anisotropy (an) $I_{c,an} = I_{c+}- I_{c-}$ measured on single loops with $w_{n} = 0.2 \ \mu m$,  $w_{w} = 0.3 \ \mu m$ and $T_{c} = 1.23 \ K$ at different temperature $T = 1.271 \ K$ (1) and $1.243 \ K$ (2). }
\end{figure}

The expected magnetic dependencies of the critical current $I_{c+}(\Phi/\Phi_{0}), I_{c-}(\Phi/\Phi_{0})$, Fig.3, allow to explain the periodical dependencies $V_{dc}(\Phi/\Phi_{0})$ of the rectified voltage observed on asymmetric loop and its system \cite{Dub2003}. But the experimental periodical dependencies $I_{c+}(\Phi/\Phi_{0}), I_{c-}(\Phi/\Phi_{0})$, Fig.3,5, differ in essence from the expected one. First of all we have found that the anisotropy dependencies $I_{c,an}(\Phi/\Phi_{0})$ correspond rather multiple $I_{c,an} \propto \overline{n} -\Phi/\Phi_{0}$ than single $I_{c,an} \propto n -\Phi/\Phi_{0}$ measurement, Fig.5. The anisotropy dependencies observed at difference temperature on different rings with $w_{n} = 0.2 \ \mu m$ and $w_{w} = 0.4; 0.3; 0.25 \ \mu m$ can be described by the relation $I_{c,an}(\Phi/\Phi_{0},T) = \overline{I_{p}}(\Phi/\Phi_{0},T) (w_{w}/w_{n} - w_{n}/w_{w})$, where $\overline{I_{p}}(\Phi/\Phi_{0},T) = I_{p}(0)(\overline{n} -\Phi/\Phi_{0})(1 - T/T_{c})$ are close to the expected values of the persistent current.

It is more strange that the magnetic dependencies of the critical current measured in opposite directions are similar $I_{c+}(\Phi/\Phi_{0}) \approx  I_{c-}(\Phi/\Phi_{0}+\Delta \phi )$ and its anisotropy $I_{c,an} = I_{c+} - I_{c-} \neq 0$ results from a shift $\Delta \phi = \Delta \Phi/\Phi_{0} \neq 0$ of these dependencies one relatively another. This shift is very strange since the minimums of the $I_{c+}(\Phi/\Phi_{0}), I_{c-}(\Phi/\Phi_{0})$ are observed at $\Phi = (n+0.5+\Delta \phi/2)\Phi_{0}$ and $\Phi = (n+0.5-\Delta \phi)\Phi_{0}$ but not at $\Phi = (n+0.5)\Phi_{0}$ as it should be expected, Fig.3, and as it is observed for symmetrical ring \cite{Mosh2005}. The shift value $\Delta \Phi $ can be measured enough precisely, up to $\pm 0.02\Phi_{0}$, because of the $I_{c+}(\Phi/\Phi_{0}), I_{c-}(\Phi/\Phi_{0})$ periodicity and the possibility to observe many, up to 25, periods.

Our measurements in the temperature interval $0.96 \div 0.99T_{c}$ have shown that with this precision the shift has the same value $\Delta \phi = 0.5$ for all asymmetric ring with $w_{w}/w_{n} \neq 1$ which we investigated. The shift value does not change with both temperature and value of ring anisotropy in the investigated interval $w_{w}/w_{n} = 2 \div 1.25$: $\Delta \phi = 0.5 \pm 0.02$ for the rings with $w_{w} = 0.4 \ \mu m$ $w_{w} = 0.3 \ \mu m$ and $w_{w} =  0.25 \ \mu m$, at $w_{n} = 0.2 \ \mu m$. We did not investigated for the present a ring with $1 < w_{w}/w_{n} < 1.25$ but it is obvious that the shift should change in this interval $w_{w}/w_{n} = 1 \div 1.25$ from zero to the maximum value $\Delta \phi = 0.5$ since it must be equal zero $\Delta \phi = 0$ for the ring with ideal symmetry. Our investigations of the rings with equal semi-ring width $w_{n} = w_{w} = 0.4 \ \mu m$ showed that it is difficult to make an enough symmetrical ring. We found that  $I_{c+}(\Phi/\Phi_{0}) = I_{c-}(\Phi/\Phi_{0})$ without a visible shift only for one of three rings whereas for two others the best coincidence $I_{c+}(\Phi/\Phi_{0}) = I_{c-}(\Phi/\Phi_{0}+\Delta \phi)$ is observed at $\Delta \phi = 0.05$ and $\Delta \phi = 0.07$.

The shift $\Delta \phi = 0.5$ observed for asymmetric ring with $w_{w}/w_{n} \geq 1.25$ can not be explained by the difference between the total magnetic flux $\Phi = BS + \Phi_{I} = BS + L_{n}I_{n} - L_{w}I_{w}$ and the flux $\Phi \approx BS$ induced only by an external magnet since it does not change with the ring asymmetry $w_{w}/w_{n}$ and when he critical current $I_{c}(T)$ changes in some times with temperature, Fig.5. The additional flux $\Phi_{I} = L_{n}I_{n} - L_{w}I_{w} = LI_{ext}(s_{w}-s_{n})/(s_{w}+s_{n}) + LI_{p}$ induced by the external current $I_{ext} = I_{c}$ and the persistent current is small $\Phi_{I} < 0.04\Phi_{0}$ in our ring, in contrast to the case of asymmetric loop with double Josephson junctions investigated in \cite{asym2Jj} (see also \cite{Barone}), because of the small inductance $L \approx 2 \ 10^{-11} \ G$ and small value of the critical current of the ring near $T_{c}$: $I_{c} < 30 \ \mu A$ at $T > 0.96T_{c}$.

\begin{figure}
\includegraphics{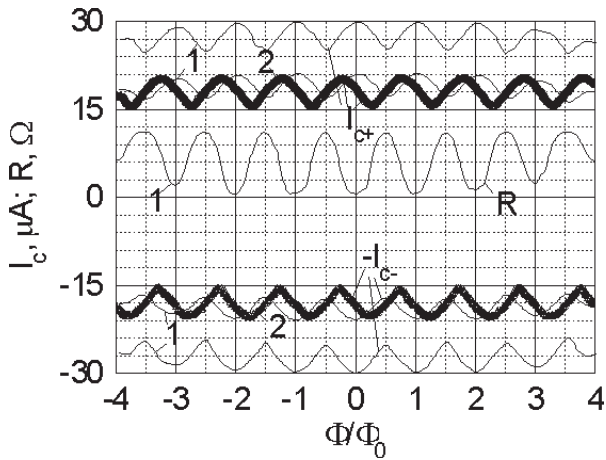}
\caption{\label{fig:epsart} The oscillations of the resistance $R$ at $T = 1.356 \ K$, of the critical current $I_{c+}$, $I_{c-}$ at $T = 1.307 \ K$ measured on symmetrical ring with $w_{n} = w_{w} = 0.4 \ \mu m$, $T_{c} = 1.358 \ K$ (1) and $I_{c+}$, $I_{c-}$ measured on asymmetric ring with $w_{n} = 0.2 \ \mu m$,  $w_{w} = 0.25 \ \mu m$, $T_{c} = 1.275 \ K$ (2) at T = 1.236 K.}
\end{figure}

It is obvious that the quantum oscillations both of the critical current and the resistance are consequence of the velocity quantization (1) and we should expect that the minimums $I_{c+}(\Phi/\Phi_{0}), I_{c-}(\Phi/\Phi_{0})$ should correspond to the maximums of the $v \propto n -\Phi/\Phi_{0}$, as well as the maximums $\Delta R(\Phi/\Phi_{0})$ correspond to the maximums of $v^{2} \propto (n -\Phi/\Phi_{0})^{2}$ both for symmetrical and asymmetric, Fig.3, rings. The observations of the minimums of $I_{c+}(\Phi/\Phi_{0}) = I_{c-}(\Phi/\Phi_{0}) = I_{c}(\Phi/\Phi_{0})$ and the maximums of $\Delta R(\Phi/\Phi_{0})$ dependencies measured on symmetrical rings at $\Phi = (n+0.5)\Phi_{0}$, Fig.6, correspond to this expectation. But the shift of the critical current dependencies measured on rings with $w_{w}/w_{n} \geq 1.25$ relatively the one measured on the symmetrical one, $I_{c+}(\Phi/\Phi_{0}) =  I_{c}(\Phi/\Phi_{0} - 0.25)$, $I_{c+}(\Phi/\Phi_{0}) =  I_{c}(\Phi/\Phi_{0} + 0.25)$ Fig.6, reveal the contradiction between the results of measurements made on the same asymmetric superconducting loop: the experimental $\Delta R(\Phi/\Phi_{0})$ dependencies, Fig.4,6,  gives evidence the maximums $v^{2}$ at $\Phi = (n+0.5)\Phi_{0}$ and $v^{2}=0$ at $\Phi = n\Phi_{0}$ both for symmetrical and asymmetric rings whereas the $I_{c+}(\Phi/\Phi_{0})$, $I_{c-}(\Phi/\Phi_{0})$ dependencies measured on asymmetric rings contradict to this conclusion. According to the principle of realism if two experimental results contradict against each other one of their is incorrect. Therefore our results are either incorrect or are experimental evidence of violation of the principle of realism on the mesoscopic level.

This work has been supported by a grant "Quantum bit on base of micro- and nano-structures with metal conductivity" of the Program "Technology Basis of New Computing Methods" of ITCS department of RAS, a grant 04-02-17068 of the  Russian Foundation of Basic Research and a grant of the Program "Low-Dimensional Quantum Structures" of the Presidium of Russian Academy of Sciences.


\begin{thebibliography}{99}

\bibitem{Legget02} A. J. Leggett, {\em J.Phys. Con. Mat.} {\bf 14}, R415 (2002); {\em J. of Supercon.} {\bf 12}, 683 (1999).

\bibitem{QuanCom} M.A.Nielsen and I.L.Chuang, {\it Quantum Computation and Quantum Information} Cambridge University Press, 2000.

\bibitem{Steane} A. M. Steane, {\it Rept.Prog.Phys.} {\bf 61}, 117, 1998.

\bibitem{EPR} A.~Einstein, B. Podolsky, and N. Rosen {\it Phys. Rev.} {\bf 47}, 777 (1935).

\bibitem{Bell} J. S. Bell, {\em Physics} {\bf 1}, 195 (1964).

\bibitem{nonlocal} A. Aspect, P. Grangier, and G. Roger, {\it Phys.Rev. Lett} {\bf 47} 460, (1981); P. G. Kwiat et al., {\em idid} {\bf 75}, 4337 (1995);  G. Weihs et al., {\em idid} {\bf 81}, 5039 (1998); W. Tittel et al., {\it Phys. Rev. A} {\bf 57}, 3229 (1998); J. W. Pan et al., {\it Nature} {\bf 403}, 515 (2000).

\bibitem{Legget85} A. J. Leggett and A. Garg {\it Phys. Rev. Lett.} {\bf 54}, 857, (1985).

\bibitem{Mermin} N.D. Mermin, {\em Physics Today}, {\bf 38}, 38 (1985).

\bibitem{tink75} M.Tinkham, {\em Introduction to Superconductivity.} McGraw-Hill Book Company (1975).

\bibitem{Little62} W.~A.~Little and R.~D.~Parks, {\em Phys.Rev.Lett.} {\bf 9}, 9 (1962).

\bibitem{Mosh92} H.~Vloeberghs et al.,  {\em Phys. Rev.Lett.} {\bf 69}, 1268, 1992.

\bibitem{Nik2001} A.V. Nikulov, {\em Phys. Rev. B} {\bf 64}, 012505 (2001).

\bibitem{Dub2003} S.~V~.Dubonos et al.,  {\em Pisma Zh.Eksp.Teor.Fiz.} {\bf 77}, 439 (2003) ({\em JETP Lett.} {\bf 77}, 371 (2003)); available also at cond-mat/0303538.

\bibitem{Mosh2005} D.~S.~Golubovic and V.~V.~Moshchalkov, will be published in {\em Appl. Phys. Lett.}, available at cond-mat/0509332.

\bibitem{asym2Jj} T. A. Fulton, L. N. Dunkleberger, and R. C. Dynes, {\em Phys. Rev. B} {\bf 6}, 855 (1972)

\bibitem{Barone} A.Barone and G.Paterno, {\em Physics of the Josephson Effect}. Wiley, New York, 1982




%\bibitem{Dub2} J. R. Friedman et al., {\it Nature} {\bf 406}, 43, (2000); C. H. van der Wal,  A. C. J. ter Haar, {\it Science} {\bf 290}, 773, (2000).

%\bibitem{Dub4} Y.~Makhlin, G.~Schoen, and A.~Shnirman, {\it Rev.Mod.Phys.} {\bf 73}, 357 (2001); P.~Bertet, et al.,  {\em Phys. Rev. Lett.} {\bf 95}, 257002 (2005); Z.~H.~Peng, M.~J.~Zhang, and D.~N.~Zheng, {\em Phys. Rev. B} {\bf 73}, 020502(R) (2006).

%\bibitem{Dub02} S.V. Dubonos, V.I.Kuznetsov and A.V. Nikulov, {\em in Proceedings of 10th International Symposium "NANOSTRUCTURES: Physics and Technology"} St Petersburg: Ioffe Institute, p. 350 (2002); physics/0105059

%\bibitem{1967} A.Th.A.M. De Waele, W.H.Kraan, R. De Bruynouboter and K.W. Taconis, {\em Physica} {\bf 37}, 114 (1967).

%\bibitem{Nik98} A.V. Nikulov and I.N. Zhilyaev, J. Low Temp.Phys. 112, 227 (1998).

%\bibitem{Feynman} R.P.Feynman, R.B.Leighton, M.Sands, {\it The Feynman Lectures on Physics}, Vol.1, Addison-Wesley Publishing Company, Reading, Massachusetts, 1963.


%\bibitem{normal} L.P.Levy et al. {\it Phys. Rev.Lett.} {\bf 64}, 2074 (1990); V.Chandrasekhar et al. {\it Phys.Rev. Lett.} {\bf 67}, 3578 (1991); E.M.Q.Jariwala et al. {\it Phys. Rev.Lett.} {\bf  86}, 1594 (2001).

%\bibitem{semicond} D.Mailly, C.Chapelier, and A.Benoit, {\em Phys.Rev.Lett.} {\bf 70}, 2020 (1993); B.Reulet et al, {\em Phys. Rev.Lett.} {\bf 75}, 124 (1995); W.Rabaud  et al, {\em Phys. Rev.Lett.} {\bf 86}, 3124 (2001).

%\bibitem{Kulik} I.O.Kulik, {\it Pisma Zh.Eksp.Teor.Fiz.} {\bf 11}, 407 (1970) ({\it JETP Lett} {\bf 11}, 275 (1970)); {\em Zh.Eksp.Teor.Fiz.} {\bf 58}, 2171 (1970).

\end{thebibliography}
\end{document}